\documentstyle[12pt]{report} 
\def\rbuildrel#1\over#2{\mathrel{\mathop{#2}\limits_{#1}}}

\catcode`@=11
\def\underline#1{\relax\ifmmode\@@underline#1\else
        $\@@underline{\hbox{#1}}$\relax\fi}

\def\titlepage{\pagestyle{empty}\c@page=0
      \def\thefootnote{\fnsymbol{footnote}} }
\def\endtitlepage{\pagestyle{plain}\c@page=1
      \def\thefootnote{\arabic{footnote}} \c@footnote\z@ }
\catcode`@=12

\newskip\humongous \humongous=0pt plus 1000pt minus 1000pt

\newif\ifdtup

\def\*{\hskip .06 cm}

\def\thebibliography#1{\section*{\ \markboth
 {REFERENCES}{REFERENCES}}\list
 {{\arabic{enumi}}.}
 {\settowidth\labelwidth{{#1}.}\leftmargin\labelwidth
 \advance\leftmargin\labelsep
 \usecounter{enumi}}
 \def\newblock{\hskip .11em plus .33em minus -.07em}
 \sloppy
 \sfcode`\.=1000\relax}

\def\thebibliographyp#1{\section*{\ \markboth
 {Chan \& Dill, $\quad$ Polymer Principles in Protein Structure
 and Stability}{Chan \& Dill, $\quad$ Polymer Principles in Protein Structure
 and Stability}}\list
 {\arabic{enumi}.}{\settowidth\labelwidth{#1.}\leftmargin\labelwidth
 \advance\leftmargin\labelsep
 \usecounter{enumi}}
 \def\newblock{\hskip .11em plus .33em minus -.07em}
 \sloppy
 \sfcode`\.=1000\relax}

\def\sqr#1#2{{\vcenter{\vbox{\hrule height.#2pt
\hbox{\vrule width.#2pt height#1pt \kern#1pt
\vrule width.#2pt}
\hrule height.#2pt}}}}

\begin{document}
%
%\titlepage
\noindent
$\null$
\hfill April 8, 2003

\vskip 0.6in 

\begin{center}
{\Large\bf Contact Order Dependent Protein Folding Rates:}\\

\vskip 0.3cm

{\Large\bf Kinetic Consequences of a Cooperative Interplay}\\ 

\vskip 0.3cm

{\Large\bf Between Favorable Nonlocal Interactions and}\\

\vskip 0.3cm

{\Large\bf Local Conformational Preferences}\\

\vskip .5in 
{\bf H\"useyin K{\footnotesize{\bf{AYA}}}}
and
{\bf Hue Sun C{\footnotesize{\bf{HAN}}}}$^\dagger$\\
$\null$

Protein Engineering Network of Centres of Excellence (PENCE),\\
Department of Biochemistry, and
Department of Medical Genetics \& Microbiology,
Faculty of Medicine, University of Toronto,
Toronto, Ontario M5S 1A8, Canada\\

%

%{\tt Submitted to ""}
%{\tt To appear in ""}
%

\end{center}

\vskip 1cm

\noindent
{\bf Running title:} Physics of Contact-Order Dependent Protein Folding \\

\vskip 1cm

\noindent {\bf Key words:} 
calorimetry / chevron plot / G\=o models /
simple two-state kinetics /\\ single-domain proteins / nonadditivity

$\null$\\ 
%

%\vskip 0.8in

\noindent 
$^\dagger$ Corresponding author.\\ 
E-mail address of Hue Sun C{\footnotesize{HAN}}: 
chan@arrhenius.med.toronto.edu\\
Tel: (416)978-2697; Fax: (416)978-8548\\
Mailing address: Department of Biochemistry, University of Toronto, 
Medical Sciences Building -- 5th Fl., 1 King's College Circle,
Toronto, Ontario M5S 1A8, Canada.

\vfill\eject 
%\endtitlepage
%%%%%%%%%%%%%%%%%%%%%%%%%%%%%%%%%%%%%%%%%%%%%%%%%%%%%
%

\def\thefootnote{\fnsymbol{footnote}}

\noindent 
{\large\bf Abstract}\\

\vskip .2 in 

\noindent 
Physical mechanisms underlying the empirical correlation between
relative contact order (CO) and folding rate among naturally-occurring 
small single-domain proteins are investigated by evaluating postulated
interaction schemes for a set of three-dimensional 27mer lattice 
protein models with 97 different CO values. Many-body interactions are 
constructed such that contact energies become more favorable when short 
chain segments sequentially adjacent to the contacting residues adopt 
native-like conformations. At a given interaction strength, this 
scheme leads to folding rates that are logarithmically well correlated 
with CO (correlation coefficient $r=0.914$) and span more than 2.5 orders of 
magnitude, whereas folding rates of the corresponding G\=o models with 
additive contact energies have much less logarithmic correlation with CO 
and span only approximately one order of magnitude. 
The present protein chain models also exhibit calorimetric cooperativity 
and linear chevron plots similar to that observed experimentally for 
proteins with apparent 
simple two-state folding/unfolding kinetics. Thus, our findings 
suggest that CO-dependent folding rates of real proteins may arise
partly from a significant positive coupling between nonlocal contact 
favorabilities and local conformational preferences.

%**************************************************************************
\vfill\eject 

$\null$
\vskip -1cm

%------------------------------------------------------------------------------

\centerline{\bf INTRODUCTION}

$\null$

\noindent
{\bf Generic protein properties as energetic constraints}

The folding of many small single-domain proteins is well approximated
by simple two-state thermodynamics and kinetics.$^{1,2}$ In
the past several years, we have shown that fundamental insights 
into protein energetics can be gained by using these general, apparently 
mundane properties as experimental constraints on protein chain 
models.$^{3-10}$ This approach is based on the recognition 
that model interaction schemes capable of producing these commonly observed 
experimental properties are, somewhat surprisingly, not entirely 
straightforward to come up with. To date, much advance has been made by 
coarse-grained modeling of protein folding.$^{7,11-15}$ Nonetheless,
the interactions postulated by many existing models are insufficient for 
calorimetric two-state cooperativity.$^{3,4}$ Furthermore, even common 
G\=o models are not cooperative enough for simple two-state kinetics, 
their explicit native biases notwithstanding. Specifically, we recently 
found that several lattice$^{6,9,10}$ and continuum (off-lattice)$^{8}$ 
G\=o-like formulations with essentially additive interaction schemes all 
led to chevron rollovers --- a hallmark of folding kinetics that are often 
operationally referred to as non-two-state.$^{9}$ Apparently, many-body 
interactions are needed to produce chevron plots with linear folding and 
unfolding arms consistent with a two-state description of equilibrium 
thermodynamics.$^{10}$ 
\\

Small single-domain proteins are characterized as well by a significant
correlation between relative contact order (CO) and folding rate.$^{16}$
Therefore, it is only logical to require a model protein interaction
scheme to produce a similar correlation.$^{17,18}$ Ising-like$^{19,20}$ and 
other$^{21,22}$ constructs without explicit chain representations have had 
successes in this regard.  However, as for thermodynamic and kinetic 
cooperativities, achieving the CO dependence requirement in models with 
explicit chain representations appears to be a nontrivial task.
Notably, an early lattice model study using a 20-letter alphabet suggested
that proteins with higher CO should fold faster,$^{23}$ thus predicting
a trend opposite$^{17}$ to that for real single-domain proteins.$^{16,18}$
A more recent 20-letter lattice model investigation, on the other hand,
found modest correlations between increasing CO and longer logarithmic 
folding time (correlation coefficient $r\approx 0.70$--$0.79$ for chain 
lengths $\ge 54$).$^{24}$ An earlier continuum G\=o model studies of 18 
proteins also found a modest correlation between increasing CO and slower 
logarithmic folding rates ($r=0.69$).$^{25}$ But the corresponding dispersion
in simulated folding rates covers only $\approx 1.5$ orders of magnitude, 
which is much narrower than the $\approx 5$ orders of magnitude covered by 
the real folding rates of the proteins in the given dataset. When a different 
potential function was used in a more recent continuum G\=o model analysis, 
however, no correlation between CO and simulated folding rates was
discerned.$^{26}$
\\

Recently, based on lattice 27mer simulations, Jewett et al.$^{27}$ have
proposed that enhanced thermodynamic cooperativity and 
many-body interactions --- which are basic properties of individual 
two-state proteins to begin with$^{1-10}$ --- may also be a key to 
understand the correlation between CO and folding rate across different 
proteins. This is an attractive and insightful idea. However, the particular 
way in which thermodynamic cooperativity was enhanced by these authors 
led only to modest increases in folding rate dispersion relative to that 
for the corresponding lattice G\=o models with pairwise additive contact 
energies. Both the dispersion in folding rates and the correlation of
logarithmic folding rate with CO ($r=0.75$) for the most cooperative 
interaction scheme they reported were
similar to that obtained from an earlier continuum G\=o model study,$^{25}$
as well as that from a recent simulation of 20-letter lattice models$^{24}$ 
with only pairwise additive contact energies (see above). In our view,
these results suggest that while CO-dependent folding may well derive from 
certain intraprotein interactions that are also responsible for high 
thermodynamic cooperativity, CO-dependent folding does not arise from 
thermodynamic cooperativity {\it per se}. In other words, how cooperativity 
is achieved can be critically important. Many {\it a priori} many-body 
mechanisms are consistent with high thermodynamic cooperativity. An example 
is the two rather different interaction schemes we considered
in ref.~10 --- one involves local-nonlocal coupling while the other
assigns an extra favorable energy to the ground-state structure
as a whole. But perhaps not all such mechanisms can mimic experimentally
observed CO dependencies to the same degree. Therefore, to shed light on 
the physical mechanisms of CO-dependent folding, we endeavor to construct
an interaction scheme that would provide larger dispersions in folding 
rates and better correlations with CO.
\\

$\null$

%------------------------------------------------------------------------------

\centerline{\bf MODELS AND METHODS}

$\null$

The present study focuses on the idea of a cooperative interplay 
between local conformational preferences and the contact-like interactions 
that drive the packing of the protein core.$^{3,5,6,10}$ We have shown
that chain models embodying this idea can lead to calorimetric
cooperativity and simple two-state kinetics,$^{10}$ although our 
exploration thus far has been limited to model proteins that are 
mostly helical.$^{3,5,6,10}$ Here we consider a general formulation
of this idea, the basic ingredients of which are described by Fig.~1A.
This hypothesis may be viewed as a synthesis of the local-dominant and 
the nonlocal-dominant perspectives.$^{28}$ We were motivated by the recognition 
that both local$^{29,30}$ and nonlocal$^{31,32}$ intraprotein interactions 
are important determinants of protein structure and stability. Yet local 
conformational preferences alone are often insufficient for stable secondary 
structures under physiological conditions. Secondary structure formation 
is known to be context dependent;$^{33}$ they are stable when packed in the 
core of a protein but are usually not stable in isolation (ref.~31 and 
references therein). Furthermore, conformational space grows exponentially 
with chain length, even when preferences arising from local excluded 
volume effects are taken into account.$^{34}$ It follows that a large 
part of the stability and uniqueness of protein native structures cannot
be explained by local interactions alone.$^{35}$ On the other hand,
our recent G\=o-model studies have shown that nonlocal contact-like
interactions by themselves are not cooperative enough for simple 
two-state kinetics$^{6,8-10}$ if they are not coupled to local
conformational propensities.
\\

\noindent
{\bf A simple model of local-nonlocal coupling}

Here we explore the hypothesis in Fig.~1A by incorporating its form of
local-nonlocal coupling into a new interaction scheme in Fig.~1B 
for explicit-chain models configured on three-dimensional simple 
cubic lattices. This allows the idea to be tested quantitatively. Fig.~1B 
may be viewed as a generalization of similar constructs we have 
used previously in the context of helical proteins.$^{3,5,6,10}$
As a first step in our inquiry, we make the simplifying assumption
that the interactions are native-centric,$^{25-27,31,36-38}$ in that
only native interactions are favored, while nonnative interactions
are neutral (have zero energy). The local-nonlocal coupling in Fig.~1B 
involves nonadditive many-body interactions. A chain segment which is 
locally nativelike (with native bond and torsion angles) but make no 
native contact is not stabilized (contributing zero energy). On the other 
hand, nonlocal contact interactions between monomers far apart along 
the chain sequence are more favorable when the chain segments around the 
contacting residues are in their native conformations than when they are not. 
As such, the present model differs from models that additively combine contact 
energies and local favorabilities.$^{39}$ The importance of nonadditive 
many-body effects in protein folding has been recognized,$^{3,5,6,10,40-44}$
but they have not been used extensively to model calorimetric two-state 
cooperativity and linear chevron plots.$^{3-10}$ Our aim here is to 
utilize extremely coarse-grained representations as a computationally
efficient means to explore the general principles linking CO-dependent 
folding and proteinlike cooperativities. Many structural and 
energetic details of real proteins are beyond the scope of this work. 
In particular, the present work does not deal with the microscopic 
physical origins of local-nonlocal coupling. Instead we just presume
that its presence in naturally occurring proteins could arise from 
evolutionary design. Because of these, the simple interaction scheme 
in Fig.~1B should be viewed only as a tentative model in this regard.
\\

In order to examine the folding rates of a set of model proteins 
whose native structures cover a diverse range of CO values, we now consider
chains of length $n=27$ configured on simple cubic lattices. For these 
27mers, there are 103,346 distinct maximally compact conformations
(not related by rotations or inversions)$^{45,46}$ confined to a 
$3\times 3\times 3$ cube. The distribution of CO among these maximally
compact conformations covers 97 different values$^{27}$ from 
CO $=$ $208/756=0.275$ to $402/756=0.532$ (inset of Fig.~2A,
where CO is computed using equation~1 of ref.~16). For each CO value, 
we randomly choose a maximally compact 27mer conformation as 
the native structure of a model protein (Table~I).\footnote{Since the 
present choices of structures are independent of that by 
Jewett et al.,$^{27}$ the structures listed in Table I do not necessarily 
coincide with those used in their study.}
\\

Folding and unfolding kinetics are modeled by standard Monte Carlo
simulations using the Metropolis criterion and the elementary
chain moves of end flips, corner flips, crankshafts, and rigid 
rotations. The relative frequencies of attempting these moves are
4.7\%, 58.3\%, 27\%, and 10\% respectively (c.f. ref.~6)\footnote{The
following typographical error in ref.~6 should be corrected. The relative 
attempt frequencies of corner flips and crankshafts used in this prior 
study of ours were, respectively, 60.6\% and 27\%, not the 27\% and 60.6\% 
stated on p.~901 of ref.~6.} 
Time is measured by the number of attempted Monte Carlo moves for a given
process. The set of elementary chain moves is chosen to mimic physically
plausible processes. Lattice model kinetics are dependent on 
the choice of move set.$^{12}$ Nonetheless, we expect the general trend 
predicted by the model is less sensitive to move set when kinetics are 
not dominated by trapping events,$^{12}$ as is the case here and has
been verified by Jewett et al.$^{27}$
Progress towards the native state is tracked by the fractional 
number of native contacts $Q$ (ref.~3--6). To ascertain the implications 
of the local-nonlocal coupling we proposed, results from a highly cooperative 
interaction scheme with $a=0.1$ are compared with that from the additive 
scheme ($a=1$) of common G\=o models (c.f. Fig.~1B). Folding trajectories 
are initiated at a randomly generated conformation; folding first passage 
time is defined by the formation of the $Q=1$ ground-state conformation.
Unfolding trajectories are initiated at the ground-state conformation;
unfolding first passage time is the time it takes for the chain to
be left with three or fewer native contacts ($Q\le 3/28$); $Q=3/28$
is chosen to define unfolding because it coorresponds approximately
to the free energy minimum for the denatured state.
\\

$\null$

%------------------------------------------------------------------------------

\centerline{\bf RESULTS}

$\null$

\noindent
{\bf Sensitivity of folding rate on CO enhanced by local-nonlocal coupling}

Fig.~2 provides the correlation between CO and folding rate among 
our 27mer models. It shows clearly that the local-nonlocal coupling
mechanism postulated in Fig.~1 can lead to a significant enhancement of
correlation as well as much increased sensitivity of folding rate to CO.
Whereas the dispersion in folding rates among the common additive 
G\=o models in Fig.~2A covers only approximately one order of magnitude
(a factor of ten) and the logarithmic folding rates exhibit only a relatively 
weak correlation with CO (correlation coefficient $r= 0.63$), the corresponding 
dispersion among the $a=0.1$ cooperative models in Fig.~2B covers 
approximately 2.5 to 3 orders of magnitude, with a strong correlation 
between CO and logarithmic folding rate ($r=0.914$) comparable to that 
observed among 
a selection of real, small single-domain proteins.$^{18}$ Similar to
the corresponding experimental situations,$^{16,18}$ the comparisons in 
Fig.~2 were performed under conditions for which folding relaxation is 
essentially single-exponential, as is evident from the good agreements
in Fig.~2 between median first passage time divided by $\ln 2$ 
and the corresponding mean first passage time.$^{6,47}$ To better
delineate the effects of having weakened contact interactions when 
the chain segments locally adjacent to the contacting residues are 
nonnative, several $a$ values other than the $a=0.1$ used for the main 
plot are compared in the inset of Fig.~2B. It shows CO-dependent
folding at different levels of local-nonlocal coupling (different 
$a$ values) for several 27mers with representative CO's. 
The $a=0$ case here corresponds to complete interdependence between 
nonlocal contact and local structure. This inset indicates
that sensitivity of folding rate to CO increases 
(the fitted line has a more negative slope) with decreasing $a$, and that
the behavior of the $a=0.1$ models is very similar to that of the
$a=0$ models. These results
further affirm that local-nonlocal coupling is a key ingredient for
the good correlation between CO and fold rate in these models. Nevertheless,
as for real proteins,$^{16,18}$ despite the good correlation, CO by itself 
cannot predict folding rates of the present models with high accuracy. 
Folding rates here can vary significantly for different structures with 
the same CO as well. For example, for the particular 27mer with CO 
$=346/756=0.458$ in Fig.~2B, the datapoint 
$\log_{10}({\rm folding\ rate})=-5.75$ may be viewed as an ``outlier'' 
vis-\`a-vis the fitted line. However, for two other 27mers with the
same CO but do not belong to the randomly chosen set in Table~I (and
therefore not plotted and not used in the correlation analysis of 
Fig.~2B), we found $\log_{10} ({\rm folding\  rate})=-7.26$ and $-7.60$, 
which happen to be much closer to the fitted line in Fig.~2B.
The reasons behind variations in folding rates among structures
with same CO remain to be elucidated.
\\

\noindent
{\bf A consistent model of thermodynamic and kinetic cooperativity}

Fig.~3 provides further analyses of the folding/unfolding kinetics of 
one example 27mer structure we choose to study in more detail.
Consistent with our previous results,$^{6,8-10}$ it shows that the model 
chevron plot$^{48}$ predicted by the common additive G\=o potential (upper 
plot) deviates significantly from simple two-state kinetics in that it 
exhibits a severe rollover under only moderately native conditions. More 
specifically, for this case rollover becomes significant at 
${\cal E}/k_{\rm B}T$ values that are only slightly more negative (more 
favorable to folding) than that of the transition midpoint 
(${\cal E}/k_{\rm B}T\approx -1.43$). In contrast, the chevron plot
predicted by the model with a substantial local-nonlocal coupling (lower 
plot) is qualitatively similar to that of real, small single-domain 
proteins that fold and unfold with simple two-state kinetics.$^{10}$
In particular, it has essentially linear folding and unfolding arms over an 
extended range of ${\cal E}/k_{\rm B}T$ values. We have also obtained 
for this model the equilibrium free energy of unfolding 
$\Delta G_{\rm u}$ as a function of ${\cal E}/k_{\rm B}T$, where 
$\Delta G_{\rm u}$ here is taken to be that between the unique $Q=1$ 
conformation and those with $Q\le 3/28$. (The same definition is used for 
unfolding kinetics as stated above.) Because $\Delta G_{\rm u}$ is 
essentially linear in ${\cal E}/k_{\rm B}T$, the linearity of the
chevron arms over an extended ${\cal E}/k_{\rm B}T$ 
range implies an essentially linear relationship between folding/unfolding 
rates and $\Delta G_{\rm u}$ within the corresponding regime (i.e., the 
model parameter ${\cal E}$ may be eliminated in favor of the lower horizontal 
scale in Fig.~3). Furthermore, comparing the mean first passage times in Fig.~3 
versus the corresponding median first passage times divided by $\ln 2$ shows 
that folding or unfolding relaxation for this model is essentially 
single exponential$^{6,47}$ for $\Delta G_{\rm u}<$ $10k_{\rm B}T$. 
Essentially single-exponential folding under moderately folding conditions
is further demonstrated by an approximately linear logarithmic 
distribution of first passage time$^{8,9,49}$ shown in the inset. 
Similar to the cooperative models we recently investigated,$^{10}$ 
for the model with local-nonlocal coupling in Fig.~3, the thermodynamic 
$\Delta G_{\rm u}$ values matches well with the kinetically obtained 
quantity $-k_{\rm B}T\ln [({\rm folding\ rate})/({\rm unfolding\ rate})]$ 
for $\Delta G_{\rm u}$ ranging from $10k_{\rm B}T$ to $-6k_{\rm B}T$
(lower V-shape). In other words, the folding/unfolding kinetics of this
model is simple two-state$^{6,8-10}$ within a $\Delta G_{\rm u}$
range quite similar to that experimentally accessible to small 
single-domain proteins.$^{10}$ Finally, the cooperative model in Fig.~3 
is also calorimetrically two-state. Assuming that the interactions are 
temperature independent, the model's van't Hoff to calorimetric enthalpy 
ratio $\Delta H_{\rm vH}/\Delta H_{\rm cal}$ ($\kappa_2$ without baseline 
subtraction$^{4}$) is determined to be $0.992$ (detailed calculation not 
shown), satisfying the requirement of 
$\Delta H_{\rm vH}/\Delta H_{\rm cal}\approx 1$ for two-state
thermodynamics.$^{3-5}$ Taken together, the above considerations imply that 
the local-nonlocal coupling mechanism for enhanced CO-dependent folding 
in Fig.~2B also provides --- as it should --- a consistent account of 
thermodynamic and kinetic cooperativities$^{6,8-10}$ in simple two-state 
proteins (Fig.~3).
\\

As it stands, the transition midpoints of all 27mers considered here 
with the local-nonlocal coupling parametrized by $a=0.1$ are very close 
to one another. This is because the interaction scheme in Fig.~1B assigns 
the same energy ($=28{\cal E}$) to every ground-state conformation. 
This is a simplifying assumption in the present modeling setup.
Since the thermodynamic stabilities of real, small single-domain proteins 
are quite diverse,$^{16,18}$ it is important to note that, in a broader
perspective, our hypothesis that significant CO-dependent folding can 
emerge from local-nonlocal coupling is not contingent upon the different 
proteins in question having very similar thermodynamic stabilities. In
more sophisticated models, for example, an extra favorable energy that 
differs from one 27mer to another may be assigned to the ground-state 
conformation (i.e., a different $E_{\rm gs}$ term as defined in ref.~10
for each 27mer). In that case, the 
thermodynamic stabilities of different 27mers can be very different, 
but their folding rates would not be affected by this extra feature of 
the model. In other words, the correlation between CO and folding rate 
in Fig.~2B would remain unchanged. As we have recently argued,$^{10}$ 
such extra stabilizing energies for the ground state as a whole are 
physical plausible because experimental evidence$^{50}$ indicates that 
in real proteins there is a partial separation between the driving 
forces for folding kinetics and the interactions responsible for
thermodynamic stability.
\\

$\null$

%------------------------------------------------------------------------------

\centerline{\bf DISCUSSION}

$\null$

Energy landscapes of the present models are further characterized in 
Fig.~4 for three representative structures with low, intermediate, 
and high CO values. In this figure, the low- and high-CO structures are, 
respectively, the fastest and slowest folding among the 97 structures 
in Table~I, whereas the intermediate-CO structure is the one analyzed
in Fig.~3. For the common additive G\=o potential, energy $E$ is
directly proportional to $Q$ ($E={\cal E}Q$). However, for the
cooperative models with local-nonlocal coupling, there are multiple 
energy levels for each $Q$, with $E={\cal E}Q$ as the lower bound
(left panels of Fig.~4). This means that, on average, the energetic 
separations between non-ground-state and ground-state conformations in the 
cooperative models with local-nonlocal coupling are larger than that in the 
additive G\=o models. This feature is demonstrated directly in the right 
panels of Fig.~4, which show that the number of non-ground-state conformations 
within a given energy range is smaller for the cooperative models than for 
the additive G\=o models except for the highest energies ($E\approx 0$).
It follows that the overall thermodynamic cooperativities of the models
with local-nonlocal coupling are substantially higher than that of the 
corresponding additive G\=o models. This behavior is expected as well from 
our recent finding that simple two-state folding/unfolding kinetics (Fig.~3
above) requires ``near-Levinthal'' thermodynamic cooperativity.$^{10}$
Indeed, for the three models in Fig.~4 with local-nonlocal coupling, 
the van't Hoff to calorimetric enthalpy ratios 
$\Delta H_{\rm vH}/\Delta H_{\rm cal}$ are, from top to bottom,
$\kappa_2=$ $0.972$, $0.992$, and $0.998$. These values are extremely
high for model enthalpy ratios without baseline subtractions.$^{4}$
In contrast, the corresponding additive G\=o models are less cooperative, 
with $\kappa_2=$ $0.751$, $0.861$, and $0.878$. Here it is noteworthy 
that the additive G\=o models' $\Delta H_{\rm vH}/\Delta H_{\rm cal}$ 
ratios even after empirical baseline subtractions,$^{4}$ 
$\kappa_2^{({\rm s})}=$ $0.885$, $0.961$, and $0.962$, are lower than 
the $\Delta H_{\rm vH}/\Delta H_{\rm cal}$ ratios of the
cooperative models in the absence of baseline subtractions.
\\

\noindent
{\bf Contact-order dependence indicative of special mechanisms 
of cooperativity}\\

Obviously, thermodynamic cooperativity is a necessary ingredient
for any protein chain model that purports to rationalize the
generic properties of small single-domain proteins.$^{3-10}$
For the particular interaction scheme we consider, the above analysis shows
that features that give rise to significant CO-dependent folding also 
lead to high thermodynamic cooperativity. However, the converse
is not necessarily true. More in-depth considerations and a comparison 
of the present results with that of Jewett et al.$^{27}$ indicate that 
higher thermodynamic cooperativity {\it per se} does not necessarily give 
rise to more enhanced dependence of folding rate on CO. Our reasoning
is as follows. First, for the present set of 27mer structures we have 
chosen randomly, the correlation between logarithmic folding rate and CO 
is quantified by $r=0.63$ ($r^2=0.39$) for the additive G\=o interaction 
scheme. Despite that this correlation happens to be weaker than that 
of Jewett et al.'s collection of additive G\=o models (their $r^2=0.51$), 
after cooperativity is enhanced by local-nonlocal coupling, the correlation 
between logarithmic folding rate and CO for our $a=0.1$ models is much 
higher ($r^2=0.84$, see Fig.~2 above, an improvement in $r^2$ value
of $0.33$)\footnote{
Because all the model chains in the present study have the same length 
and the same number of native contacts, their correlation coefficient 
between folding rate and CO is the same as that between folding rate 
and the total contact distance (TCD) defined in ref.~51.
}
than the best case reported by Jewett et al.$^{27}$ ($r^2=0.57$ 
for their $s=3$, an improvement in $r^2$
value of $0.06$ over that for their additive G\=o models).\footnote{
If the $s=3$ interaction scheme of Jewett et al. is applied to the
present set of structures and kinetic models, we found $r^2=0.65$ 
for the correlation between CO and folding rate. In
this case, the folding rates span $\approx 1.8$ orders of magnitude; 
see ref.~52 for details.
}
Second, the folding rates of our cooperative models are much more
sensitive to CO, covering 2.5 to 3 orders of magnitude, whereas those
of Jewett et al. cover only approximately 1.3 orders of magnitude.
This means that the present local-nonlocal coupling mechanism is 
significantly more effective in enhancing CO dependence than the
nonlinear $E$--$Q$ relationship postulated by Jewett et al. (equation~1
of ref.~27). Their interaction scheme does not make direct reference to 
chain conformations as such. Thermodynamic cooperativity is
enhanced in their models by stipulating that the total contact energy $E$ (for
a given conformation as a whole) does not decrease (does not become more 
favorable) linearly with increasing $Q$ as in common G\=o models; 
but rather decreases at progressively faster and faster rates when $Q$ 
is closer to unity.\footnote{
Jewett et al. suggested that the ``extraordinary
cooperativity in protein folding'' may originate from ``three-body 
interactions.'' But how three-body interactions might lead to their 
$E$--$Q$ relationship remains to be elucidated.} 
Third, in fact, if thermodynamic cooperativity is further increased
in the interaction scheme of Jewett et al. by increasing their $s$ parameter, 
the energy landscape will eventually become a Levinthal golf course in the 
$s\rightarrow\infty$ limit. In that case, folding would be rate-limited 
by random conformational search and CO-dependence would be all but 
eliminated. Fourth, in this connection, we have recently considered three 
27mer models with CO $=0.28$, $0.40$ and $0.51$ in a separate study. The 
thermodynamic cooperativity of these models are enhanced by
assigning an extra stabilizing energy to the ground state but without 
local-nonlocal coupling.$^{10}$ For the energetic parameters we considered,
the folding rates of these models cover less than an order of 
magnitude.$^{10}$ The same set of results also indicated that dispersion 
in folding rates under moderately folding conditions would decrease if 
thermodynamic cooperativity is increased by assigning an even stronger 
stabilizing energy to the ground state, in a manner similar to greatly
increasing $s$ in Jewett et al.'s formulation. Taken together, these 
observations lead us to the conclusion that while thermodynamic 
cooperativity is certainly necessary, by itself it is not sufficient 
to guarantee CO-dependent folding rates similar to that observed 
experimentally$^{16,18}$ if the underlying mechanism for thermodynamic
cooperativity is not specified.
\\

CO-dependent folding highlights the important role of local interactions 
in determining folding rates.$^{16-18}$ It suggests that the mechanism
of folding may involve relatively fast formation of local structure.
In this regard, we note that under the general lattice scheme in Fig.~1B,
formation of strong (unattenuated) native contacts with contact order 
$\vert j - i \vert=3$ is relatively easier than formation of strong
native contacts with higher contact orders. This is because in the
$\vert j - i \vert=3$ case there is an overlap between parts of the two 
local segments that have to be nativelike in order 
for the contact to be strong. Physically, how a general 
mechanism similar to that in Fig.~1 may arise in real proteins from 
solvent-mediated atomic interactions such as sidechain packing
and hydrogen bonding remains to be elucidated. 
Many basic issues will have to be tackled to address this question. 
For example, correlations between backbone and sidechain rotamer 
conformations$^{53}$ may contribute to such a mechanism.
Another possibility is that aspects of {\it anti-cooperativity} 
of certain types of hydrophobic interactions$^{54}$ may help disfavor 
premature nonspecific hydrophobic collapse (which would lead to kinetic 
trapping$^{14}$) when the sidechains are locally less well packed than 
that in the native state. If this is the case, it could give rise to 
local-nonlocal coupling mechanisms similar to that postulated in Fig.~1.
\\

In summary, while the models used in the present study are rudimentary,
they provide strong evidence that a cooperative interplay between local 
conformational preferences and nonlocal favorable contact-like
interactions is an important mechanism in accounting for experimentally
observed CO-dependent folding of small single-domain proteins.
We are optimistic that more rigorous applications of
the CO-dependence constraint as well as the thermodynamic and kinetic 
cooperativity requirements would help further narrow down theoretical 
possibilities and thus contribute to a more realistic understanding of protein 
energetics. 
\\

%%%%%%%%%%%%%%%%%%%%%%%%%%%%%%%%%%%%%%%%%%%%%%%%%%%%%%%%%%%%%%%%%%%%%%%%%%%

%---------------------------------------------------------------------------

$\null$

%===========================================================================

\noindent
{\Large Acknowledgments.}
We thank Robert L. Baldwin, Alan Davidson, Teresa Head-Gordon, Michael 
Levitt, Vijay Pande, Kevin Plaxco, Steve Plotkin, Wes Stites and Yaoqi Zhou 
for helpful discussions, and Vijay Pande and Kevin Plaxco for kindly sharing 
their work (ref.~27) before publication. The research reported 
here was partially supported by the Canadian Institutes of Health 
Research (CIHR grant no. MOP-15323), a Premier's Research Excellence Award 
from the Province of Ontario, and the Ontario Centre for Genomic Computing 
at the Hospital for Sick Children in Toronto. H. S. C. is a Canada 
Research Chair in Biochemistry.

%===========================================================================
\vfill\eject

\par\vfill\eject

\noindent
{\large\bf References}

\kern -1.5cm

%------------------------------------------------------------------------
\vfill\eject

\centerline{\large \bf Table I}
\vskip .2 in

{\footnotesize

\begin{center}
\begin{tabular}{|cc|cc|}
\hline
$\sum\Delta S_{ij}$ & conformation & $\sum\Delta S_{ij}$ & conformation \\
\hline
 208 & uufddfuurddbuubddruufddfuu  & 306 & uufrrbldrfdflurullddburdbr \\
 210 & uufddfuurbbdffdbbrffuubdbu  & 308 & uufdfrbrbulddrfllfrruublfl \\
 212 & ufdfuubbrddffubufrddbuubdd  & 310 & ufrulblfddrrbbllfuburdrfub \\
 214 & uuffdbdfrbufubbddrffuubdbu  & 312 & uufrbbllffdrrdllbubdrurfdb \\
 216 & ufdfuubbrddfuufddruubbddfu  & 314 & uufdrubbdfdfllbbuffubbrddr \\
 218 & ufdfuubbrdfufddbbruuffddbu  & 316 & uffrddblbruufdllbdffrulubb \\
 220 & uuffddburfdbbuuffrddbuubdd  & 318 & uufrbddbuullffdrrdllbubdru \\
 222 & uuffddburdfuubbddruuffdbdf  & 320 & uufddfrruubbdfdbluuffldrbd \\
 224 & uufddrbufubrfdbdfflurulldd  & 322 & uffdbrbrfufullbbrdrufldfdr \\
 226 & uufddfuurddbubdrffubbulfrf  & 324 & ufrbddlfrflurullbbddffubrr \\
 228 & uffdbrbuffdrbbuffubbllfrfl  & 326 & ufrfddluulddbbuufdrdbrfubu \\
 230 & ufdrbufublffddruurddbuubdd  & 328 & ufrubdbuldldrrffllbufurblb \\
 232 & uufrrblddrufdluldfurdruull  & 330 & uufdrubblddlfubuffddrrbbuf \\
 234 & uufddfuurddrbluurfdbbulddr  & 332 & uuffrddruubbdfllfdbrrbluuf \\
 236 & ufddbbuurrflfrdlbdfrbubldr  & 334 & ufrrdbdfluldbbrruuflbldrfd \\
 238 & uuffdbdfrrblbrulffrulbbrfd  & 336 & uffrbrbuflblffrrddllbrrblu \\
 240 & uffdbrfurdbblufrbuffllbbrf  & 338 & uffurrdldrbblurullfrrdldlf \\
 242 & uufrbddfflburflurrbbdffdbb  & 340 & ufrrbbdffdlbrbllffurbubldr \\
 244 & ufdrrbluulfrdrbuffllddrrul  & 342 & ufrullbrrblldrrdllfufdrrbu \\
 246 & ufdfurbdfruullbrrddblurull  & 344 & uffdrdllbrbluuffdbrrdbuuff \\
 248 & ufdfrbuflurblbrrdldrffuubd  & 346 & uuffddrrbbuufdfuldbubddflu \\
 250 & ufddbrfruublfdbrdblluurdru  & 348 & uufrbbdlulddrrffllbuufdrrb \\
 252 & uffdbrrflurbbdlufufrbbllff  & 350 & ufubrrdfdfuldblfuurrbldbdr \\
 254 & ufddbrblurrdfflubrfulbrbll  & 352 & uffddrbllurrfubbddlluuffdd \\
 256 & ufdfurdruullbbrddrfluurdbu  & 354 & ufrfdrbufubbllfrflddbrburd \\
 258 & ufrbdflfrrbbuullfrrdfulldr  & 356 & uffrrbdbuullffrrbldbdflfrr \\
 260 & uufddfurbbrdlffrbufubblffl  & 358 & uufdrrubddffuulldrdlbrbuuf \\
 262 & uuffdbrbufrfldrdllbrbrfubu  & 360 & ufrddllfrruulldrblubddrruu \\
 264 & ufdrurddbuuldblurrddllffrb  & 362 & ufubrrdffldrbblflfuurrbldb \\
 266 & ufdrurddllbrbluurrfldbrdfu  & 364 & ufrfddlbblffubbuffrdbrdbuu \\
 268 & uuffdbrubrfddbuldflfrrulur  & 366 & ufrfdbdfllbbuufdfurdbdbruu \\
 270 & uuffrrdllbdrbufrulbrddffll  & 368 & uffurrbbddffllbrbuulfrdrfl \\
 272 & uufrdfuldbdfrruubblddfrubd  & 370 & uffurrddbbuufllbrddflfrubr \\
 274 & ufdrubrfddllbbuurrdldrfuld  & 372 & uufdrfdruubbddluufflddbrru \\
 276 & ufdrbdlfrrubdblluurffrbbdl  & 374 & uffrdrbbuullffrrdbuldbdflf \\
\end{tabular}
\end{center}
\vskip .15 in

}

\vfill

$\null$ \hfill $\dots$ {\it Table I to be cont'd}

\vfill\eject

\noindent{\large \bf Table I} $\dots$ ({\it cont'd from last page})
\vskip .2 in

{\footnotesize

\begin{center}
\begin{tabular}{|cc|cc|}
 278 & ufddrrbllbrrullurrfflbdfrb  & 376 & uffrddllbuubddrfrbuufdlflu \\
 280 & uffrddlubdruubddllfubuffdd  & 378 & ufdrfdlluubbdfdbrfrbuuffld \\
 282 & ufrbdffuldlubbddrrfflbuldf  & 380 & ufrfddlbrbllffubbuffrdbrbu \\
 284 & uufrfldrrubbldrfdblfuldfrr  & 382 & uufdrfdrbbuufdfullddbrbuuf \\
 286 & uffubbrddrffuldlbrurbufflb  & 384 & ufrbbullddfuurrfllddrrbblu \\
 288 & ufdfrrubufldlubbrfdbdfrbuu  & 386 & uffddrbllfuubbddrruuffdbll \\
 290 & uufrbbldrfdbllfubuffddrurd  & 388 & uffrburbddffllbrbuulffrrdb \\
 292 & uffrddbbuufdldblffrulubbdf  & 390 & ufrufddrbbuffubbllffddbrbu \\
 294 & ufdrdfulurbbddlluufddrfluu  & 392 & ufrufddrbbuffubbllffddbrbu \\
 296 & ufdfrbdflbbruuffllddbbuufd  & 394 & ufrrddlbburuflblddffurbrdb \\
 298 & ufrrdblblurfrbddffluldbrbl  & 396 & uffrddblflbufubbddrruufdlf \\
 300 & ufdfrullddbuubddrffrbuubdd  & 400 & ufrfddllubdrrblluuffrdbrbu \\
 302 & ufdfurddlluubbdfdbrfrbuufd  & 402 & ufrufrbbddffllbrbuulffdrrb \\
 304 & ufrdlluurrbbdfdbllfuubdruf  &  &  \\
\hline
\end{tabular}
\end{center}
\vskip .15 in

}
{\noindent {{\bf Table~I.}}} $\quad$
The ground-state 27mer conformations ($n=27$) used in this investigation are 
given by sequences of 26 bond directions, where r = right ($+x$), 
l = left ($-x$), f = forward ($+y$), b = backward ($-y$), u = up ($+z$), 
d = down ($-z$). A structure is randomly selected for each of the 97 possible 
CO values amongst the compact 27mer structures with 
$t_{\rm max}=28$ contacts. Each integer $\sum\Delta S_{ij}$ is
the sum of $\vert j-i \vert$ over the $(i,j)$ nearest-neighbor 
contacts in the given conformation ($j-i\ge 3$). 
Here CO $=\sum\Delta S_{ij}/(n t_{\rm max})$ $=\sum\Delta S_{ij}/756$.

%------------------------------------------------------------------------
\vfill\eject

\noindent
{\large\bf Figure Captions}\\

\noindent
{\bf Figure 1.} $\quad$
(A) Schematics of local-nonlocal cooperative energetics in protein
folding. The conformation in the solid box represents the native (N) 
structure; the two filled circles depict a pair of nonlocal residues 
interacting favorably in the native state. The interaction strength 
between a residue pair is strong and essentially the same as that in 
the native structure if the chain segments sequentially local
to both residues are nativelike, as in (i). [Dotted boxes in (A) are 
used to mark nativelike chain segments.] However, the interaction strength is 
weakened if one or two chain segments sequentially local to the
interacting residues are not nativelike, as in examples (ii)--(iv). 
(B) A lattice implementation of this protein folding scenario. Here
the favorable energy for every contact (between residues $i$ and $j$, 
$\vert j-i \vert \ge 3$) in the ground-state native (N) structure is 
${\cal E}$ ($<0$) when the relative positions of the five residues centered 
at $i$ (residues $i-2$, $i-1$, $i$, $i+1$, and $i+2$) as well as the relative 
positions of five residues centered at $j$ (residues $j-2$, $j-1$, 
$j$, $j+1$, and $j+2$) are the same as that in N [solid lines in (i)],
irrespective of the relative orientations of the two five-residue chain 
segments.  However, if the local conformation of one or both sets of 
five contiguous residues is nonnative, the contact energy is weakened by an 
attentuation factor $a$ ($0\le a< 1$). Examples of the latter situation 
is given by (ii)--(iv), where nonnative local chain segments are 
drawn as broken lines.
\\

\noindent
{\bf Figure 2.} $\quad$
Correlation between the common (base 10) logarithm of folding rate 
and CO for the 97 structures in Table~I under moderately folding
conditions at ${\cal E}/k_{\rm B}T=-1.47$, using (A) the
common additive G\=o potential and (B) the local-nonlocal
cooperative interaction scheme with $a=0.1$. Solid lines are least-square
fits. Here folding rate is the reciprocal of mean folding first passage 
time (folding rate $=$ 1/MFPT).  Each MFPT is averaged from 500 
trajectories. Associated with each value of $\log_{10}(1/{\rm MFPT})$ 
(filled circle) is an open circle marking the common logarithm of the 
median folding first passage time (FPT) divided by $\ln 2$. If the 
kinetics is single-exponential, MFPT $=$ (median FPT)/$\ln 2$. 
The inset in (A) is the distribution of CO among the 
103,346 maximally compact 27mer conformations, wherein the number of 
conformations (vertical scale) is shown as a function of CO (horizontal 
scale). The inset in (B) uses six 
representative structures with different CO values ($\sum\Delta S_{ij}=$ 
208, 224, 268, 310, 348, and 386 entries in Table~I) to illustrate that 
$\log_{10}({\rm folding\ rate})$ (vertical scale) is more sensitive to CO 
(horizontal scale) when the local-nonlocal coupling is stronger. In this
inset, different symbols denote different $a$ values; the lines fitted
through the symbol are, from top to bottom, for $a=1$, $0.75$, $0.5$, 
$0.25$, $0.1$, and $0.0$.
\\

\noindent
{\bf Figure 3.} $\quad$
Model chevron plots for a CO $=0.410$ structure ($\sum\Delta S_{ij}=310$ 
entry in Table~I) are given by negative natural logarithm of MFPT as a 
function of ${\cal E}/k_{\rm B}T$ (filled symbols). Values 
of (median FPT)/$\ln 2$ are shown by the open symbols. Squares (folding)
and triangles (unfolding) are for the additive G\=o potential 
($a=1$, upper plot), whereas circles (folding) and diamonds (unfolding) 
are for the $a=0.1$ local-nonlocal cooperative interaction scheme (lower 
plot). Each MFPT is averaged from 500 trajectories, except for the model 
with local-nonlocal coupling at ${\cal E}/k_{\rm B}T=-1.47$ (arrow). 
For this particular case, 7,500 folding trajectories were simulated to 
provide enriched statistics for the FPT distribution in the inset,
wherein $P(t)\Delta t$ is the fraction of trajectories with 
$t-\Delta t/2<$ FPT $\le t+\Delta t/2$, and the bin size $\Delta t$ for FPT 
is equal to $5\times 10^6$. The free energy of unfolding 
$\Delta G_{\rm u}$ for the $a=0.1$ cooperative model is computed using
Monte Carlo histogram techniques based on sampling at the transition 
midpoint ${\cal E}/k_{\rm B}T=-1.33$. $\Delta G_{\rm u}$ is essentially
linear in ${\cal E}$ (lower horizontal scale). The dotted V-shape,
which fits well to the kinetic datapoints of the $a=0.1$ cooperative
model over an extended regime, is an hypothetical simple two-state 
chevron plot consistent with the dependence of $\Delta G_{\rm u}$ on
${\cal E}$. 
\\

\noindent
{\bf Figure 4.} $\quad$
Energy landscapes of three representative models with local-nonlocal
coupling ($a=0.1$, $\sum\Delta S_{ij}=224$, $310$, and $386$ entries 
in Table~I; ${\cal E}=-1$). The left panels show the correlation between 
$E$ and $Q$; each dot indicates that at least one conformation with the 
given $(E,Q)$ was encountered in our sampling. The right panels show these 
structures' logarithmic densities of states, where $g(E)$ is the number of 
conformations with energy $E$ for the cooperative models ($a=0.1$, dots). 
Included for comparison are the $\ln g(E)$ values of the corresponding 
additive G\=o models ($a=1$, open circles; ${\cal E}=-1$). The densities of 
states here are estimated by Monte Carlo 
sampling at the models' transition midpoints ${\cal E}/k_{\rm B}T=-1.33$ 
($a=0.1$) and ${\cal E}/k_{\rm B}T=-1.43$ ($a=1$). Note that the cooperative 
models have more energy levels than the additive models. Therefore, to 
compare their densities of states on an equal footing, the open squares 
provide the natural logarithm of the number of conformations in the $a=0.1$ 
cooperative models with energies in the range $m-0.5\le E< m+0.5$, 
where $m=1,0,-1,-2,\dots$ is an integer. Now the densities of states
represented by the open squares ($a=0.1$) are directly comparable to
that represented by the open circles ($a=1$) because their values
are based upon the same unity bin size for $E$.

%===========================================================================


\begin{thebibliography}{99}

%XXXXXXXXXXXXXXXXXXXXXXXXXXXXXXXXXXXXXXXXXXXXXXXXXXXXXXXXXXXXXXXXXXXXXXXXXXX

\bibitem{1}
Jackson SE, Fersht AR. 
Folding of chymotrypsin inhibitor 2. 1. Evidence for a two-state
transition. Biochemistry 1991;30:10428--10435.

\bibitem{2}
Baker D.
A surprising simplicity to protein folding. Nature 2000;405:39--42.

\bibitem{3} 
Chan HS.
Modeling protein density of states: Additive hydrophobic
effects are insufficient for calorimetric two-state cooperativity.
Proteins 2000;40:543--571. 

\bibitem{4}
Kaya H, Chan HS.
Polymer principles of protein calorimetric
two-state cooperativity. 
Proteins 2000;40:637--661 [Erratum: Proteins 2001;43:523].

\bibitem{5}
Kaya H, Chan HS.
Energetic components of cooperative protein folding. 
Phys Rev Lett 2000;85:4823--4826.

\bibitem{6}
Kaya H, Chan HS.
Towards a consistent modeling of protein thermodynamic and kinetic
cooperativity: How applicable is the transition state picture to
folding and unfolding? J Mol Biol 2002;315:899--909.

\bibitem{7}
Chan HS, Kaya H, Shimizu S. Computational methods 
for protein folding: scaling a hierarchy of complexities. 
In: Jiang T, Xu Y, Zhang MQ, editors. Current Topics in Computational 
Molecular Biology. Cambridge, MA: The MIT Press; 2002. p 403--447.

\bibitem{8}
Kaya H, Chan HS.
Solvation effects and driving forces for protein thermodynamic and
kinetic cooperativity: How adequate is native-centric topological
modeling? J Mol Biol 2003;326:911--931.

\bibitem{9}
Kaya H, Chan HS.
Origins of chevron rollovers in non-two-state protein folding kinetics.
Submitted (2003); [cond-mat/0302305,\\
{\tt http://xxx.lanl.gov/abs/cond-mat/0302305}].

\bibitem{10}
Kaya H, Chan HS.
Simple two-state protein folding kinetics requires near-Levinthal
thermodynamic cooperativity. Submitted (2003);
[cond-mat/0302306, 
{\tt http://xxx.lanl.gov/abs/cond-mat/0302306}].

\bibitem{11}
Bryngelson JD, Onuchic JN, Socci ND, Wolynes PG. 
Funnels, pathways, and the energy landscape of protein folding: A
synthesis. Proteins 1995;21:167--195.

\bibitem{12}
Dill KA, Bromberg S, Yue K, Fiebig KM, Yee DP, Thomas PD,
Chan HS. 
Principles of protein folding --- A perspective from simple
exact models. Protein Sci. 1995;4:561--602.

\bibitem{13}
Thirumalai D, Woodson SA. Kinetics of folding of proteins
and RNA. Acc Chem Res 1996;29:433--439.

\bibitem{14}
Chan HS, Dill KA. Protein folding in the landscape 
perspective: Chevron plots and non-Arrhenius kinetics. 
Proteins 1998;30:2--33.

\bibitem{15}
Mirny L, Shakhnovich E. 
Protein folding theory: From lattice to all-atom models.
Annu Rev Biophys Biomol Struct 2001;30:361--396. 

\bibitem{16}
Plaxco KW, Simons KT, Baker D. 
Contact order, transition state placement and the refolding rates
of single domain proteins. J Mol Biol 1998;227:985--994.

\bibitem{17}
Chan HS. 
Matching speed and locality. Nature 1998;392:761--763. 

\bibitem{18}
Plaxco KW, Simons KT, Ruczinski I, Baker D. (2000). Topology,
stability, sequence, and length: Defining the determinants of two-state
protein folding kinetics. Biochemistry 2000;39:11177--11183.

\bibitem{19}
Alm E, Baker D. Prediction of protein-folding mechanisms from
free-energy landscapes derived from native structures. 
Proc Natl Acad Sci USA 1999;96:11305--11310.

\bibitem{20}
Mu{\~n}oz V, Eaton WA.
A simple model for calculating the kinetics of
protein folding from three-dimensional structures.
Proc Natl Acad Sci USA 1999;96:11311--11316.

\bibitem{21}
Debe DA, Goddard WA. 
First principles prediction of protein folding
rates. J Mol Biol 1999;294:619--625.   

\bibitem{22}
Makarov DE, Keller CA, Plaxco KW, Metiu H. 
How the folding rate constant of simple, single-domain proteins depends
on the number of native contacts.      
Proc Natl Acad Sci USA 2002;99:3535--3539.

\bibitem{23}
Abkevich VI, Gutin AM, Shakhnovich EI.
Impact of local and nonlocal interactions on thermodynamics and kinetics
of protein folding. J Mol Biol 1995:252:460--471.

\bibitem{24}
Faisca PFN, Ball RC. Topological complexity, contact order, and
protein folding rates. J Chem Phys 2002;117:8587--8591.

\bibitem{25} 
Koga N, Takada S. 
Roles of native topology and chain-length scaling in protein folding: 
A simulation study with a G\=o-like model.
J Mol Biol 2001;313:171--180.

\bibitem{26} 
Cieplak M, Hoang TX. 
Universality classes in folding times of proteins.
Biophys J 2003;84:475--488.

\bibitem{27} 
Jewett AI, Pande VS, Plaxco KW. 
Cooperativity, smooth energy landscapes and the origins of
topology-dependent protein folding rates. J Mol Biol 2003;326:247--253.

\bibitem{28}
Uversky VN, Fink AL.
The chicken-egg scenario of protein folding revisited.
FEBS Lett 2002;515:79--83.

\bibitem{29}
Baldwin RL, Rose GD.
Is protein folding hierarchic? I. Local structure and peptide
folding. Trends Biochem Sci 1999;24:26--33.

\bibitem{30}    
Shortle D. Composites of local structure propensities:
Evidence for local encoding of
long-range structure. Protein Sci 2002;11:18--26.

\bibitem{31}
G\=o N, Taketomi H.
Respective roles of short- and long-range interactions in protein folding.
Proc Natl Acad Sci USA 1978;75:559--563.

\bibitem{32}
Dill KA.
Dominant forces in protein folding.
Biochemistry 1990;29:7133--7155.

\bibitem{33}
Minor DL, Kim PS.
Context-dependent secondary structure formation of a designed protein 
sequence. Nature 1996;380:730--734.

\bibitem{34}
Feldman HJ, Hogue CWV.
Probabilistic sampling of protein conformations: New hope for brute force?
Proteins 2002;46:8--23.

\bibitem{35}
Shimizu S, Chan HS.
Origins of protein denatured state compactness and hydrophobic 
clustering in aqueous urea: Inferences from nonpolar potentials of
mean force. Proteins 2002;49:560--566.

\bibitem{36}
Micheletti C, Banavar JR, Maritan A, Seno F. 
Protein structures and optimal folding from a geometrical variational  
principle. Phys Rev Lett 1999;82:3372--3375.

\bibitem{37}
Clementi C, Nymeyer H, Onuchic JN. Topological and energetic 
factors: What determines the structural details of the transition state
ensemble and ``en-route'' intermediates for protein folding? An investigation
for small globular proteins. J Mol Biol 2000;298:937--953.

\bibitem{38}
Linhananta A, Zhou Y. 
The role of sidechain packing and native contact interactions in folding:
Discontinuous molecular dynamics folding simulations of an all-atom
G\=o model of fragment B of {\it Staphylococcal} protein A.
J Chem Phys 2002;117:8983--8995.

\bibitem{39}
Thomas PD, Dill KA.
Local and nonlocal interactions in globular proteins and
mechanisms of alcohol denaturation.
Protein Sci 1993;2:2050--2065.

\bibitem{40}
Kolinski A, Galazka W, Skolnick J. 
On the origin of the cooperativity of protein folding: Implications
from model simulations. Proteins 1996;26:271--287.

\bibitem{41}
Plotkin SS, Wang J, Wolynes PG.
Statistical mechanics of a correlated energy landscape model for protein
folding funnels. J Chem Phys 1997;106:2932--2948.

\bibitem{42}
Liwo A, Kazmierkiewicz R, Czaplewski C, Groth M, Oldziej S, Wawak RJ, 
Rackovsky S, Pincus MR, Scheraga HA. United-residue force field for 
off-lattice protein structure simulations: III. Origin of backbone 
hydrogen-bonding cooperativity in united-residue potentials.
J Comput Chem 1998;19:259--276.

\bibitem{43}
Takada S, Luthey-Schulten Z, Wolynes PG.
Folding dynamics with nonadditive forces: A simulation study
of a designed helical protein and a random heteropolymer.
J Chem Phys 1999;110:11616--11629.

\bibitem{44}
Eastwood MP, Wolynes PG.
Role of explicitly cooperative interactions in protein folding funnels:
A simulation study. J Chem Phys 2001;114:4702--4716.

\bibitem{45}
Chan HS, Dill KA.
The effect of internal constraints on the configurations of chain
molecules. J Chem Phys 1990;92:3118--3135 [Erratum: J Chem Phys
1997;107:10353].

\bibitem{46}
Chan HS, Bornberg-Bauer E.
Perspectives on protein evolution from simple exact models.
Applied Bioinformatics 2002;1:121-144.

\bibitem{47}
Gutin A, Sali A, Abkevich V, Karplus M, Shakhnovich EI. 
Temperature dependence of the folding rate in a simple protein model:
Search for a ``glass'' transition. 
J Chem Phys 1998;108:6466--6483.

\bibitem{48}
Matthews CR. 
Effect of point mutations on the folding of globular proteins.
Methods Enzymol 1987;154:498--511.

\bibitem{49} 
Abkevich VI, Gutin AM, Shakhnovich EI. Free energy
landscape for protein folding kinetics: Intermediates, traps, and multiple
pathways in theory and lattice model simulations.
J Chem Phys 1994;101:6052--6062.

\bibitem{50}
Northey JGB, Di Nardo AA, Davidson AR.
Hydrophobic core packing in the SH3 domain folding transition state.
Nature Struct Biol 2002;9:126--130.

\bibitem{51}
Zhou HY, Zhou YQ. Folding rate prediction using total contact distance.
Biophys J 2002;82:458--463.

\bibitem{52}
Chan HS, Shimizu S, Kaya H. Cooperativity principles in protein folding.
Methods Enzymol, in press.

\bibitem{53}
Dunbrack RL. Rotamer libraries in the 21st century.
Curr Opin Struct Biol 2002;12:431--440.

\bibitem{54}
Shimizu S, Chan HS. Anti-cooperativity and cooperativity
in hydrophobic interactions: Three-body free energy landscapes and
comparison with implicit-solvent potential functions for proteins.
Proteins 2002;48:15--30 [Erratum: Proteins 2002;49:294].


\end{thebibliography}
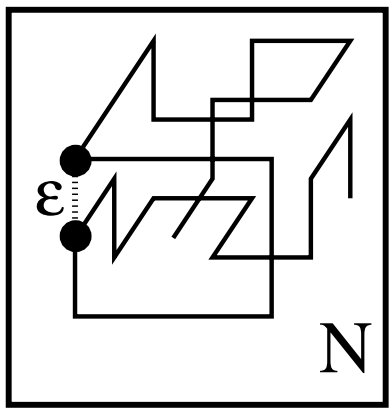
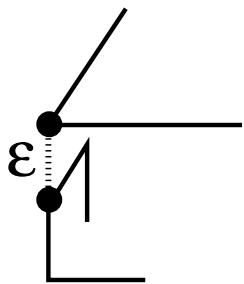
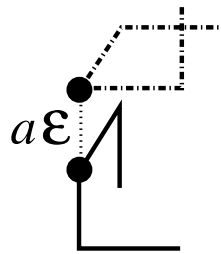
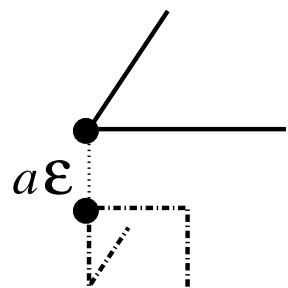
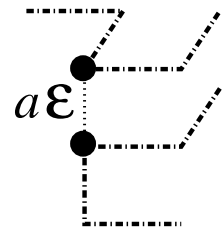
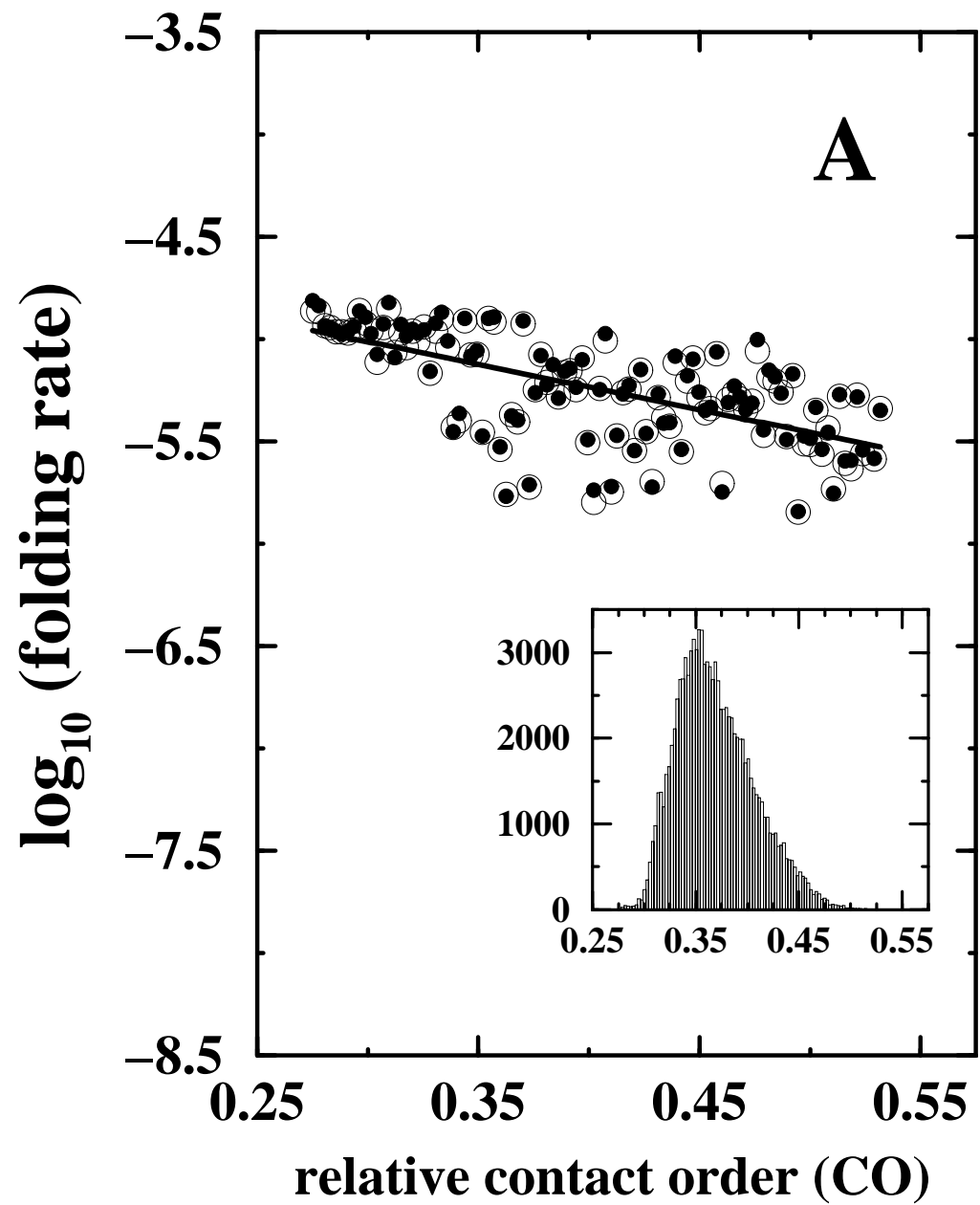
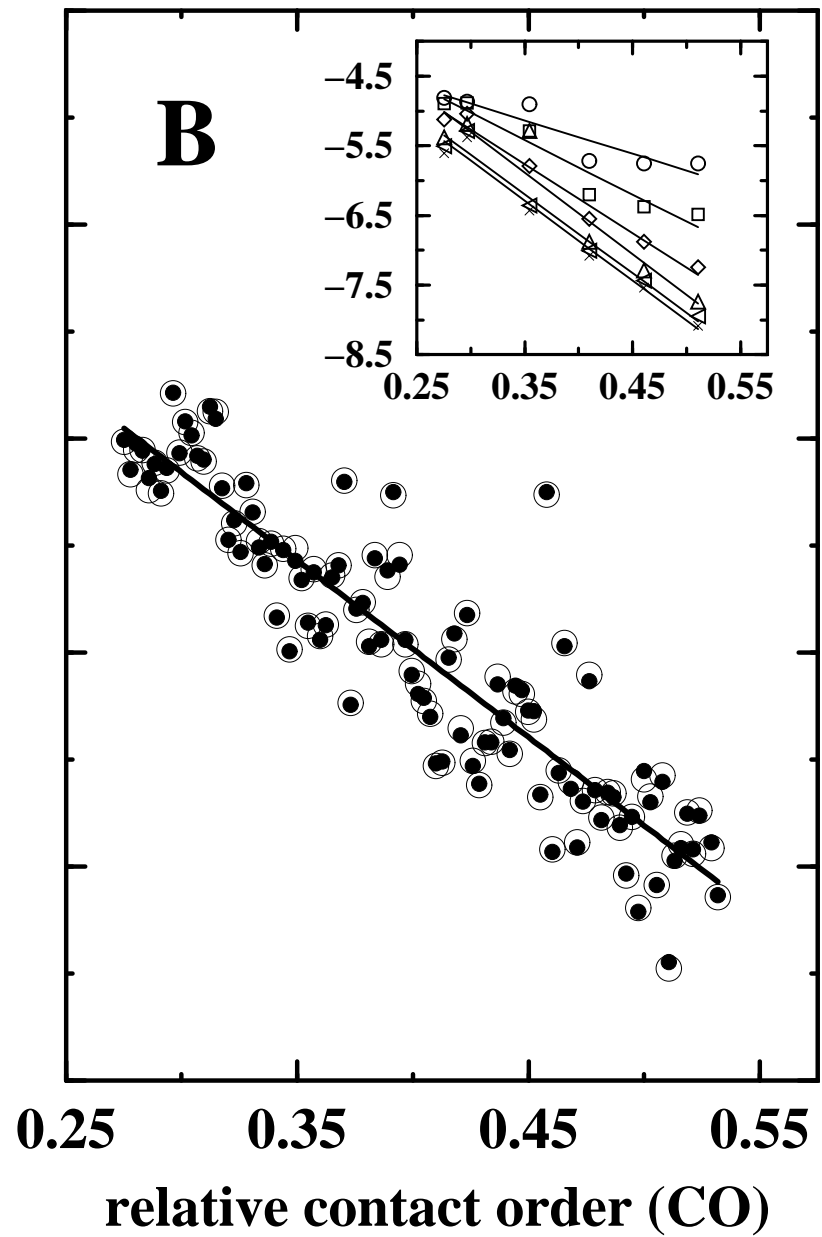
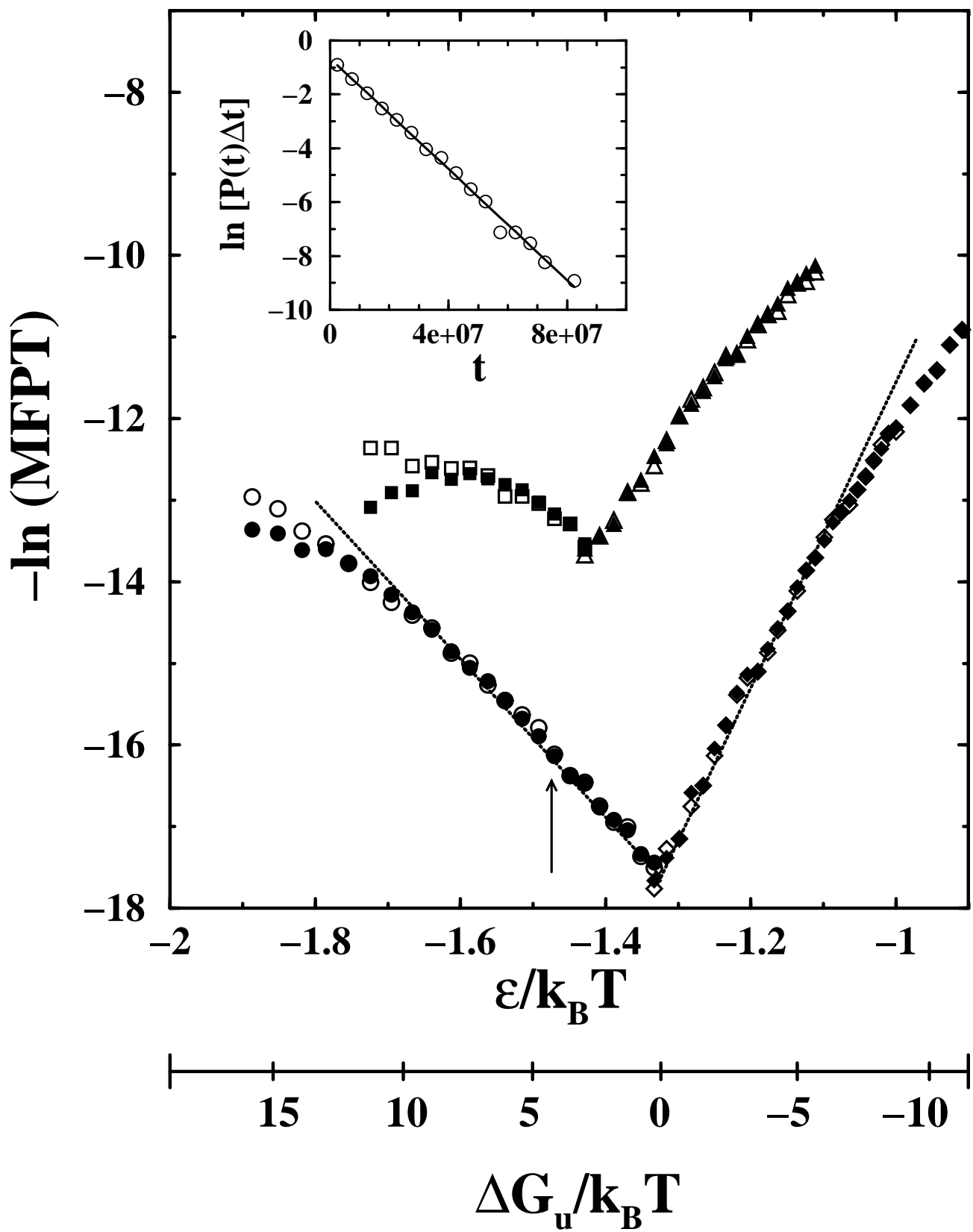
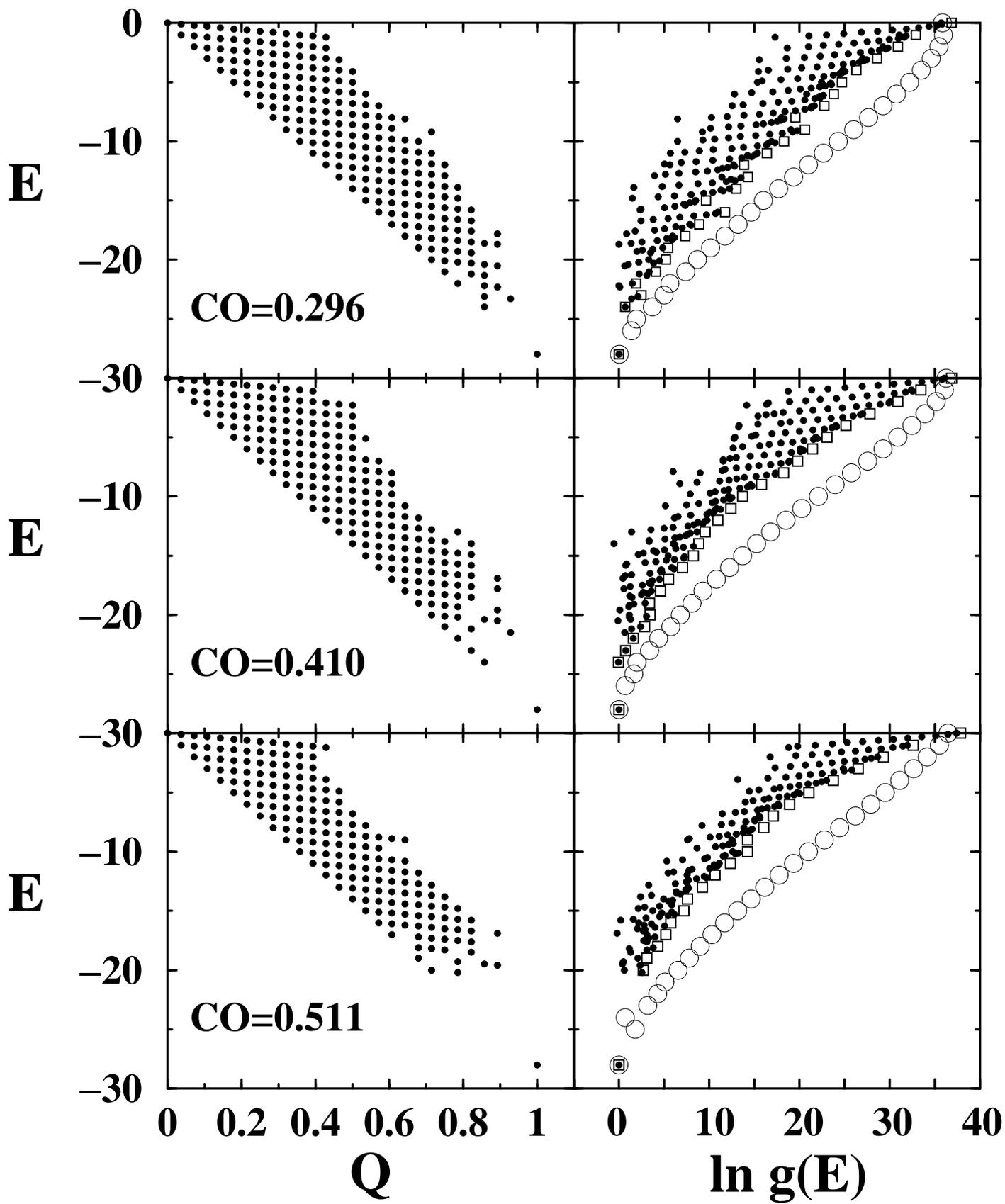
\end{document}